\documentclass[sigconf, screen]{acmart}
\usepackage{hyperref}
\usepackage{subcaption}
\usepackage{graphicx}
\usepackage{listings}
\usepackage{xcolor}
\AtBeginDocument{%
  }

\copyrightyear{2024}
\acmYear{2024}
\setcopyright{acmlicensed}\acmConference[CIKM IRS 2024]{Proceedings of the 3rd International Workshop on Industrial Recommendation Systems}{October 25, 2024}{Boise, ID, USA}
\acmBooktitle{Proceedings of the 3rd International Workshop on Industrial Recommendation Systems (IRS 2024), Held in conjunction with CIKM-2024, October 25, 2024, Boise, ID, USA}
\acmDOI{10.1145/3627673.3679083}
\acmISBN{979-8-4007-0436-9/24/10}

\begin{document}

\newcommand{\srijan}[1]{\textcolor{blue}{[Srijan: #1]}}
\newcommand{\rish}[1]{\textcolor{red}{[Rishabh: #1]}}
\newcommand{\mohit}[1]{\textcolor{orange}{[Mohit: #1]}}

\title{Crafting Tomorrow: The Influence of Design Choices on Fresh Content in Social Media Recommendation}


\author{Srijan Saket$^*$}
\email{srijanskt@gmail.com}
\affiliation{
    \institution{ShareChat}
    % \city{Seattle}
    \country{USA}
}

\author{Mohit Agarwal$^*$}
\email{mohitagarwal0212@gmail.com}
\affiliation{%
  \institution{ShareChat}
  \country{India}
}

% \author{Kartikey Pohani}
% \email{kartikeypohani@sharechat.co}
% \affiliation{%
%   \institution{ShareChat}
%   \country{India}
% }

\author{Rishabh Mehrotra}
\email{erishabh@gmail.com}
\affiliation{%
  \institution{Sourcegraph}
  \country{UK}
}

\renewcommand{\shortauthors}{Saket et al.}

\begin{abstract}
  The rise in popularity of social media platforms, has resulted in millions of new, content pieces being created every day. This surge in content creation underscores the need to pay attention to our design choices as they can greatly impact how long content remains relevant. In today's landscape where regularly recommending new content is crucial, particularly in the absence of detailed information, a variety of factors such as UI features, algorithms and system settings contribute to shaping the journey of content across the platform. While previous research has focused on how new content affects users' experiences, this study takes a different approach by analyzing these decisions considering the content itself. 
  
  Through a series of carefully crafted experiments we explore how seemingly small decisions can influence the longevity of content, measured by metrics like Content Progression (CVP) and Content Survival (CSR). We also emphasize the importance of recognizing the stages that content goes through underscoring the need to tailor strategies for each stage as a one size fits all approach may not be effective. Additionally we argue for a departure from traditional experimental setups in the study of content lifecycles, to avoid potential misunderstandings while proposing advanced techniques, to achieve greater precision and accuracy in the evaluation process.
\end{abstract}


% \ccsdesc[300]{Computer systems organization~Data flow architectures}


\received{15 August 2024}

\maketitle
\def\thefootnote{*}\footnotetext{Equal Contribution}

\section{Introduction}
% In the realm of online marketplaces the conversation, fairness has mainly focused on user-centric perspectives \cite{beutel2019fairness,geyik2019fairness,burke2017multisided,zhu2018fairness,zhu2020measuring}. Nonetheless this paper asserts that content fairness also plays a role in the operation of these platforms. More than just ensuring an experience for creators, fairness has a significant influence on aspects of the ecosystem. It plays a vital part in enhancing the success of creators motivating them to continue creating content, providing them with an opportunity, to enhance their work based on user inputs. Moreover, the impacts of content fairness extend to the realm of user engagement creating effects that ripple throughout the community. Engaged users don't just consume more content but also derive value from the platform by discovering new content and expanding their knowledge. Although there has been research on fairness \cite{mehrotra2018towards, pessach2022review, chen2023bias, ge2021towards, zhu2021fairness}, this paper seeks to explore how design choices influence the different stages of a contents life cycle highlighting its implications within online content marketplaces; Fig.\ref{fig:intro-pic}.

In the past few years, video platforms have become extremely popular, changing how people watch and interact with digital content. Platforms like Instagram Reels, TikTok, and YouTube Shorts offer a wide variety of videos that differ in length, genre, and type, making it tricky to recommend the right content effectively. ShareChat alone receives approximately 2 million new pieces of content daily, complicating efforts to determine the right UI to surface content, select the correct algorithms for accurate recommendations, and make system choices, especially when considering budget constraints.

\begin{figure}
    \centering
    \includegraphics[width=1.0\linewidth]{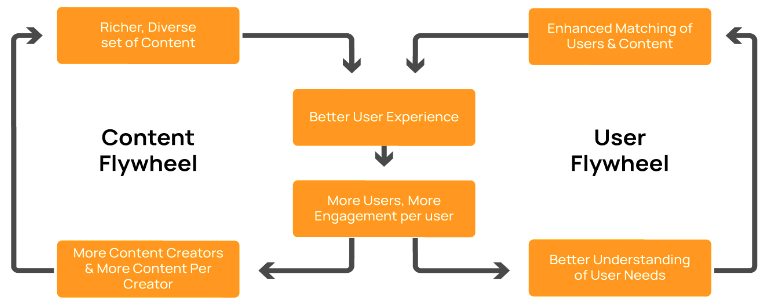}
    \caption{Content and User Flywheel}
    \label{fig:flywheel}
\end{figure}

There is a rich body of literature that discusses fairness, primarily from user-centric perspectives (either individual or group-level) \cite{ekstrand2022fairness, wang2023survey, ma2020group}. While we acknowledge the importance of this aspect in recommendation systems, we argue that it is not sufficient on its own. Extensive work has also been done on provider-side fairness and its impacts, such as Biega et al.’s work on Amortized Attention \cite{biega2018equity}, Diaz et al.’s work on Expected Exposure \cite{diaz2020evaluating}, and Zehlike et al.’s work on fair ranking \cite{zehlike2022fairness, zehlike2022fairness2}, among others. Additionally, there is further research on balancing multi-sided concerns, such as Mehrotra et al. \cite{mehrotra2018towards}, and mixed-perspective work like that of Epps-Darling et al. \cite{epps2020artist}. This paper, however, asserts that understanding the different phases content goes through and the decisions that determine content fairness also play a crucial role in the operation of these video platforms.

The content flywheel in Fig \ref{fig:flywheel} illustrates that a rich and diverse content set leads to a better user experience, which in turn drives higher user engagement, motivating creators to produce more content. The user flywheel is tightly coupled with the content flywheel in the sense that an improved user experience (driven through content and others) leads to increased engagement, providing more feedback data to better understand user interests, which ultimately results in richer training data for machine learning models. To the best of our knowledge, there is limited work on understanding the collective impact of UI, algorithm and system choices for early stage recommendation on different stages of content lifecycle [Section \ref{content_lifecycle}].

We posit that the content lifecycle in short video applications (and other content marketplaces) can be characterized by multiple distinct stages: the initial phase, the intermediate phase, and the mature phase till it finally expires \cite{plc_wu2018beyond, plc_yu2015lifecyle, plc_zhou2021survey}. Each of these stages presents unique attributes which are covered in detail in section \ref{content_lifecycle}.
% \begin{itemize}
%   \item In the \textbf{early phase}, behavioral data is notably absent, with the primary focus being on understanding the content itself.
%   \item As content progresses to the \textbf{intermediate phase}, there is a marginal increase in behavioral data available, although still coupled with a limited understanding of the content.
%   \item In the \textbf{mature phase}, an abundance of behavioral data becomes accessible for analysis and implementation of supervised machine learning models.
% \end{itemize}

These varying stages necessitate practitioners to adopt distinct models and make specific design choices tailored to the specific phase of content development \cite{plc_wang2023fresh, guo2020survey, zhang2019deep}. The decisions made regarding design at each stage of content development exert a substantial influence on the success of content within the platform. Consequently, this has a ripple effect on both users and creators, ultimately playing a crucial role in the patterns of content consumption throughout the entire platform.

This work centers on examining how design decisions made in the initial phases of content development have a cascading effect on the success of content, user experiences, and the overall consumption patterns on a large-scale platform \cite{chaney2018algorithmic, chen2021exploration, hu2019vizml}. This includes decisions including number of views received by a fresh content, algorithm used for recommendation, as well as user interface and surface choices to provide visibility to the content. 

The paper delves into various crucial aspects, showcasing how a combination of system, algorithm and design choices can have a catastrophic impact on content success within the platform. Major contributions of the paper include:
\begin{enumerate}
    \item Identifying the impact of varied minimum exposure to fresh content on content progression outcomes.
    \item Recognising the importance of differential treatment to time sensitive categories in early phase of the recommendation funnel \cite{plc_wang2023fresh}.
    \item Establishing that both the initial impression count and its rate are pivotal in shaping content longevity on the platform.
    \item Analyzing how advanced algorithms for personalisation affect content development and persistence.
    \item Examining the influence of presenting new content across various user interface options using metrics related to content advancement and user contentment.
\end{enumerate}

\section{Content Lifecycle}\label{content_lifecycle}

\subsection{Overview of Content Lifecycle}

The content lifecycle on a social media platform represents the sequential process through which content progresses, aiming to maximize its reach and engagement \cite{raza2022news, falk2019practical, wang2021survey}. This content propagation creates a flow of information where each stage informs the next, allowing the platform to adapt to user behavior and preferences over time, as shown in Fig. \ref{fig:contentlifecycle}. This ongoing process helps maintain a dynamic and engaging content ecosystem. Here’s how it works in four stages:

\begin{figure}[htb]
    \centering
    \includegraphics[width=0.45\textwidth]{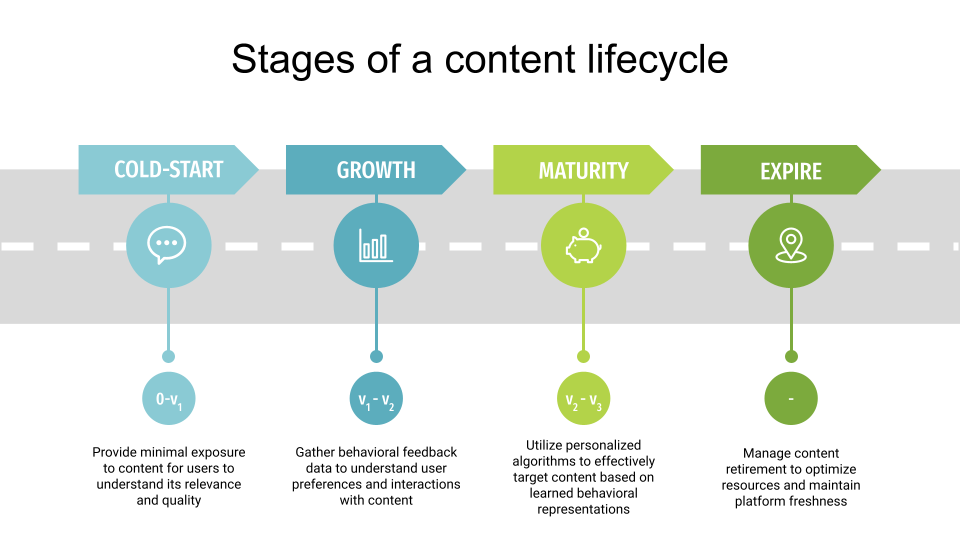}
    \caption{Proposed stages a content goes through during it's existence on a short-video platform}
    \label{fig:contentlifecycle}
\end{figure}

\begin{enumerate}
  \item \textbf{Early stage recommendation}: In the ‘early’ phase, behavioral data is notably absent. The main objective at this stage is to introduce content to a limited audience to gather initial feedback and engagement, which helps gauge its relevance and quality. This process aids in assessing the content’s potential for wider appeal.
  \item \textbf{Growth}: As content progresses to the ‘growth’ phase, there is a marginal increase in available behavioral data, although it is still accompanied by a limited understanding of the content. The key objective at this stage is to generate behavioral feedback data to learn behavioral representations \cite{zhao2015connecting, saket2024monitoring}. We analyze user interactions (likes, comments, shares, etc.) to identify patterns and preferences. This data is then used to refine the content’s targeting and presentation.
  \item \textbf{Maturity}: In the ‘mature’ phase, an abundance of behavioral data becomes available for analysis and the implementation of supervised machine learning models. The key objective at this stage is to utilize personalized algorithms for effective targeting. These personalized algorithms use the learned behavioral representations to fine-tune content recommendations, maximizing the likelihood of engagement by presenting content that aligns with user preferences.
  \item \textbf{Expiration}: Content is allowed to naturally phase out after it has run its course. Additionally, a defined threshold ensures that storage and serving resources are optimised, maintaining platform freshness.
\end{enumerate}

\subsection{Data Context}
% \srijan{TODO: Add details; improve writing}
\label{sec:data}
\subsubsection{App Description}
We consider production traffic from one of the largest multi-lingual short video applications, ShareChat \footnote{\href{https://sharechat.com/}{https://sharechat.com/}}, that delivers content in over 14 languages, with a user base exceeding 180 million monthly active users. All the content are user-generated belonging to a variety of genres. 

% We randomly sampled user interaction data over the course of two months from over 4 million users, capturing more than 14 million video posts. This data includes both implicit signals, such as Video Play—indicating the successful completion of a recommended video beyond a specified threshold—and skips, as well as explicit signals, such as likes, shares, downloads, and clicks.

\subsubsection{Details of Data}
\label{data_details}
We compiled a dataset of video posts created in the Hindi language for an in-depth analysis. We expect other languages to behave similarly because of the randomness in the data selection process. This dataset includes \textit{randomly} sampled user interaction data collected over two months from more than 4 million users, covering over 14 million video posts. This allows us to track the evolution of content across different viewership levels and its effect on user satisfaction metrics. We assessed this impact through both implicit indicators, such as successful video plays $(1 \text{ if } watch\_time > 0.98 \times video\_duration, 0 \text{ otherwise})$ and skips $(watch\_time < \textit{3s})$, as well as explicit signals like clicks, likes, and shares. Additionally, the dataset includes information about the feed that led to an impression [Section \ref{ui_surface}]. The dataset is structured as time-series data, with each post associated with a view counter at different points in time. In the experiments, users are randomly assigned to different treatment groups to assess the influence on satisfaction metrics.

\begin{figure}[htb]
    \centering
    \includegraphics[width=8cm, height=5cm]{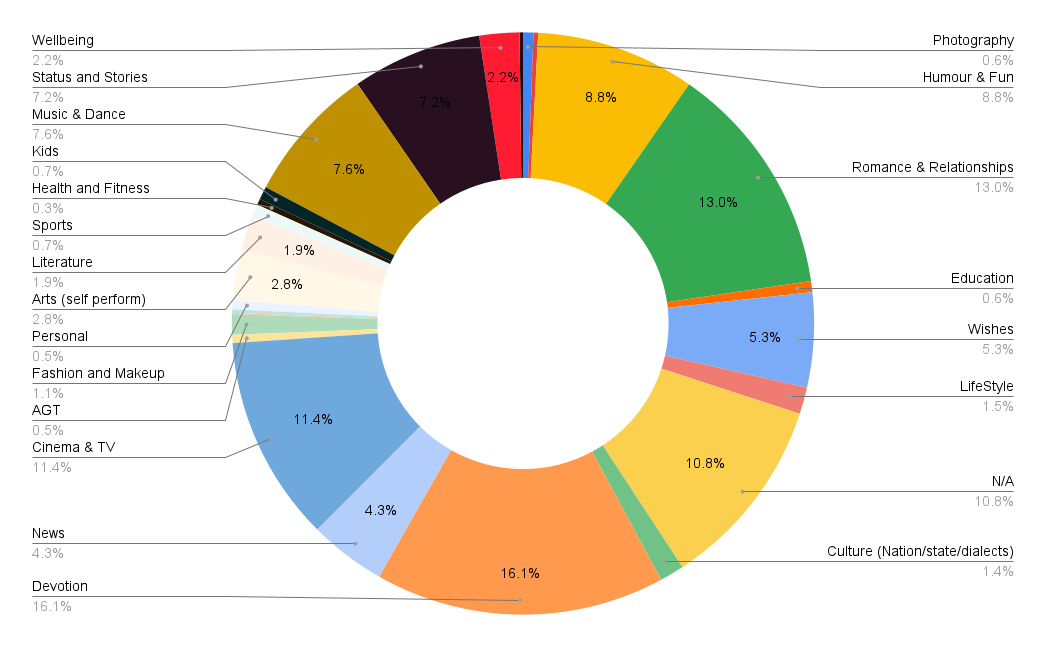}
    \caption{Genre Distribution}
    \label{fig:genre_dist}
    \vspace{-5pt}
\end{figure}

\subsubsection{Genres}
The dataset includes content that falls into about 25 broad categories, ranging from topics like \textit{Devotion} and \textit{News} to \textit{Romance \& Relationships}, among others. Some categories contain a higher density of content compared to others, as shown in Fig. \ref{fig:genre_dist}. Our analysis explores how the initial exposure of content influences changes in its distribution across these categories on the platform. Additionally, we examine whether all categories behave uniformly or if there are notable differences worth highlighting.

% \subsubsection{Users}
% We examine ShareChat application users, with a particular emphasis on those who use Hindi. In the experiments, users are randomly assigned to different treatment groups to assess the influence on satisfaction metrics.

\subsection{Content Lifecycle}
% The trajectory of a content on social media hinges significantly on its treatment in the early stages. Initial exposure serves as a litmus test for the content's quality and relevance to the user base. Misplaced targeting during this phase can detrimentally impact the content's longevity. However, this is only the beginning. Subsequent stages must be adept at generating high-quality feedback data from user interactions, discerning between exceptional and mediocre content. The ultimate aim of a robust content lifecycle is to learn a refined representation of content, amalgamating behavioral and semantic information. This culmination is what we term the "Mature" stage. The wider the proportion of this mature corpus, the broader the pool of high-quality content available for recommendation, ensuring both user satisfaction and content health. 

The success of content on short-video apps largely depends on how it’s managed in the early stages. Initial exposure acts as a test to see if the content is relevant and appealing to users. Poor targeting during this time can harm the content’s long-term success. But this is just the beginning. Later stages should focus on gathering useful feedback from user interactions to differentiate between great and average content. The goal of a strong \textit{content lifecycle} is to develop a clear understanding of the content by combining user behavior with content semantics. This final phase is known as the ‘Mature’ stage. The more content that reaches this mature stage, the larger the pool of high-quality content available for recommendations (referred to as the candidate pool \cite{davidson2010youtube, covington2016deep}), ensuring users are satisfied and the content ecosystem remains healthy.

% The success of content on short-video applications largely depends on how it’s handled in the early stages. The initial exposure acts as a test to see if the content is relevant and appealing to users. Poor targeting during this time can hurt the content’s long-term success. But this is just the start. Later stages need to focus on gathering useful feedback from user interactions to distinguish between great and average content. The goal of a strong \textit{content lifecycle} is to develop a clear understanding of the content by combining user behavior and content semantics. This final phase is known as the “Mature” stage. The more content that reaches this mature stage, the larger the pool of high-quality content available for recommendations (referred as the candidate pool \cite{davidson2010youtube, covington2016deep}), ensuring users are happy and the content ecosystem remains healthy.

\subsubsection{Definitions}
The content lifecycle unfolds in distinct stages:
\begin{itemize}
  \item \textbf{Reaching `minViews'}: At this initial stage, a fixed budget is allocated to all platform content. We represent it in this paper as \(\textit{views}_\textit{min}\) for the platform. Specific slots on a user's feed are reserved for exposure. A user's feed serves as the virtual space where content consumption takes place.

  \item \textbf{Reaching Maturity}: Achieving maturity involves learning a high-quality, low-dimensional representation of content (32 dimensions in our case), primarily through behavioral feedback gathered during initial exposure. Such systems leverage models like Field-aware Factorization Machines (FFMs), wide-and-deep neural networks, and two-tower models to learn vector representations for users and content, often referred to as \emph{embeddings} \cite{cheng2016wide, juan2016field, guo2017deepfm}. Notably, content performance metrics show improvement with refined representations \cite{saket2024monitoring}. The paper uses the average across all posts as a threshold for this stage.

\begin{figure}[htb]
    \centering
    \includegraphics[width=8cm, height=5cm]{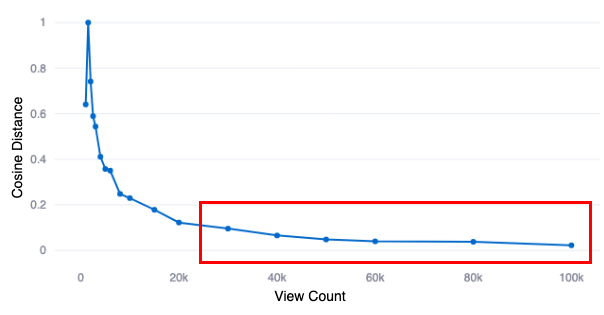}
    \caption{Embedding Maturity Curve}
    \label{fig:emb_evolution}
\end{figure}
  
  \item \textbf{Content Progression}: Content shelf-life depends on the interaction of various lifecycle stages [\ref{content_lifecycle}]. This section explores how specific design, algorithmic, and system choices affect the likelihood of content surpassing certain view thresholds. These thresholds are carefully selected to reflect peaks in the content maturity curve. Fig \ref{fig:emb_evolution} illustrates the saturating cosine distance curve of content embedding at a given time $t$ and the embedding at the converged state, plotted against the number of views $x$ the content has received on the platform. We will use \textbf{Conditional View Probability (CVP)} as one of the metrics to measure content progression in early-stage recommendation problems. CVP measures the probability of content reaching a certain number of views (x) given that it has already received at least a minimum number of views (y). We propose the following formula for CVP:
  
  % Content shelf-life hinges on the interplay of various lifecycle stages. This phase delves into the impact of specific design, algorithmic, and system choices on the likelihood of content surpassing specific view thresholds. These thresholds are strategically selected to mirror peaks in the content maturity curve, as shown in Fig \ref{fig:emb_evolution}.
  % % \srijan{cite maturity curve}\srijan{Mention the specific thresholds?}. 
  % We will use \textbf{Conditional View Probability (CVP)} as one of the metrics to measure content progression in early stage recommendation problems. CVP measures the probability of a content reaching a certain number of views (x) given that it has already received at least a minimum number of views (y). It can be calculated as:

  \begin{equation}\label{CVP}
     \text{CVP}(x \mid y) = \frac{\left|\{c \in C \mid v(c) \geq x, v(c) \geq y\}\right|}{\left|\{c \in C \mid v(c) \geq y\}\right|}
  \end{equation}

   where, \( C \) is the set of all content; \( v(c) \) is the number of views for content \( c \); \( x \) and \( y \) are view thresholds.

% \[CVP(x|y) = \frac{\textit{Number \!of \!content \!with \!at \!least } y \textit{ views that reach } x \textit{ views}}{\textit{Total number of content with at least } y \textit{ views}}\]
  
  \item \textbf{Content Success Metrics}: Evaluating content success depends on both user satisfaction and content progression metrics. Additionally, the study examines how category reach changes based on different algorithmic and system choices. We measure this using the \textbf{Content Survival Rate (CSR)}, which is the probability that a piece of content remains active and relevant over a specified period, given design choices $\tilde{X}$. By ‘active,’ we mean that the content continues to receive a significant number of views organically. We propose the following formula for CSR:

  \begin{equation}\label{CSR}
      \text{CSR}(t'\mid y, t, \tilde{X}) = \frac{\left| \{ c \in C \mid v(c, t + t') - v(c, t) \geq x \} \right|}{\left| \{ c \in C \mid v(c, t) \geq y \} \right|} \quad
  \end{equation}

  where, \textit{x,t > 0}; the numerator is the set of all items that receive a minimum of \textit{x} views in between time \textit{t} and \textit{t'}, suggesting content activeness; and the denominator is set of all items with at least \textit{y} views within time \textit{t}. 
% \[CSR(t'|y) = \frac{\textit{Number \!of \!pieces \!of \!content \!that \!remain \!active \!at \!time \!t+t'}}{\textit{Total number of contents with at least y views within time t}}\]
  
\end{itemize}

Each stage of the content lifecycle plays a crucial role in determining the overall success and longevity of content on the platform.

\begin{figure}[htb]
    \centering
    \includegraphics[width=9cm, height=5cm]{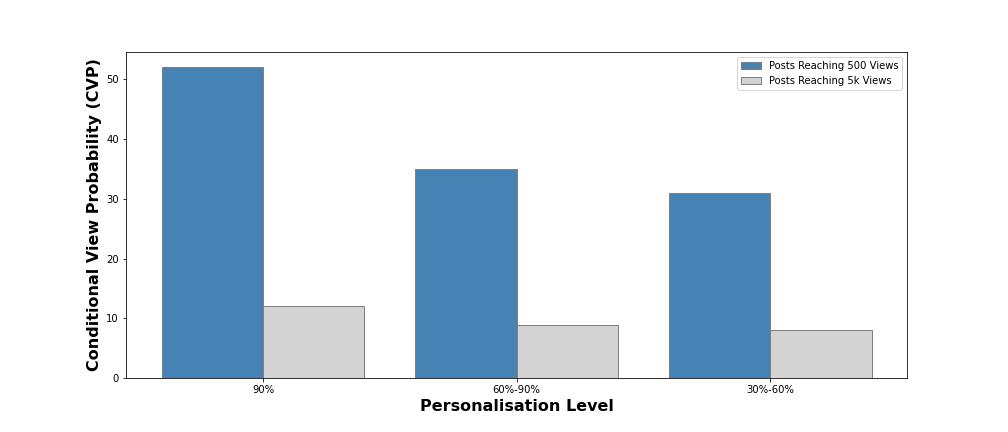}
    \caption{Effect of varying Personalisation on CVP}
    \label{fig:personalize}
\end{figure}

\begin{figure*}[t!]
    \centering
    \begin{subfigure}[t]{0.34\textwidth}
        \centering
        \includegraphics[width=\textwidth]{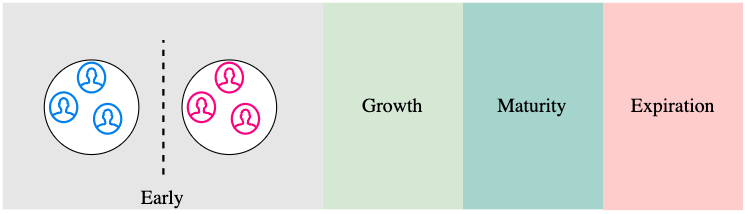}
        \caption{User level AB}
        \label{fig:challenge_1}
    \end{subfigure}
    % \hspace{0.001mm} % Add horizontal space for separation
    \vrule width 0.05pt % Vertical line as a separator
    % \hspace{0.001mm} % Add horizontal space for separation
    \begin{subfigure}[t]{0.32\textwidth}
        \centering
        \includegraphics[width=\textwidth]{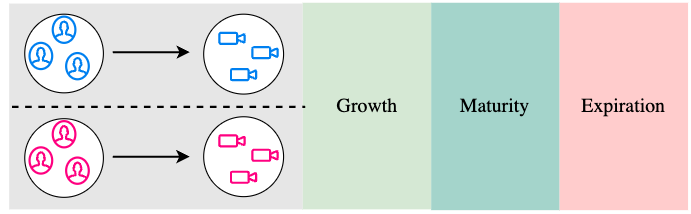}
        \caption{User-Content AB}
        \label{fig:challenge_2}
    \end{subfigure}
    % \hspace{0.001mm} % Add horizontal space for separation
    \vrule width 0.05pt % Vertical line as a separator
    % \hspace{0.001mm} % Add horizontal space for separation
    \begin{subfigure}[t]{0.3\textwidth}
        \centering
        \includegraphics[width=\textwidth]{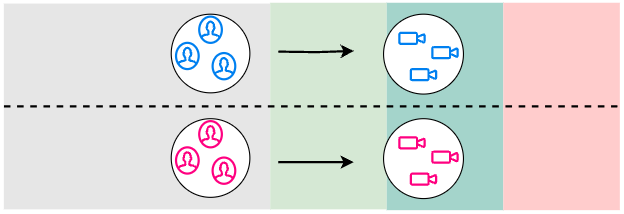}
        \caption{Parallel Experimentation Setup}
        \label{fig:challenge_3}
    \end{subfigure}
    \caption{The figure shows evolution of experimentation framework to gauge the impact of design choices across user satisfaction and platform content distribution}
    \label{fig:experimentation}
\end{figure*}

\section{Design Choices Impact Success Outcomes}
\subsection{System Design Choice}
% \begin{enumerate}
\subsubsection{Number of minimum impressions}: The value of $views_{min}$ itself is a fundamental aspect of system design choice. To investigate this, we experimented with varying $views_{min}$ while keeping the other parameters consistent. 

\subsubsection{Latency to provide minimal exposure}: Latency, defined as the time taken to ensure the post receives the minimum guaranteed views, $views_{min}$, is another significant design choice. We segmented content into different categories to understand the impact of latency on time-sensitive vs non time-sensitive categories. 

Our hypothesis is that by adjusting the $views_{min}$ based on system design choices, the platform can better improve content progression and success metrics. Additionally, optimizing latency, especially for time-sensitive content, can greatly increase the reach and effectiveness of such posts on the platform.

% Our hypothesis is that the design choice of minimizing latency for time sensitive genres  e.g. news should be more critical.
% For time sensitive genres e.g. news, minimizing latency is critical for achieving the desired view count. Our analysis demonstrated that when the $views_{min}$ were provided early in the post lifecycle, the post's tendency to reach the target view count was notably higher. In contrast, for non-time sensitive genres e.g. humor, the impact of latency on reaching the desired view count was less pronounced. Posts in these genres demonstrated a more consistent tendency to achieve the target view count irrespective of the timing of the $views_{min}$. 
    
\subsection{Algorithmic Personalisation Choices}
\label{algo_choices}
We discuss three distinct approaches regarding how personalisation in the early-stage can affect the content journey.

\subsubsection{Random allocation} Each new content is assigned a random 32-dim embedding at the start, devoid of any personalized information or guidance from the platform. This approach provides no prior knowledge about user preferences or content genres. This method is simplest to implement but lacks personalisation, potentially resulting in suboptimal recommendations. Users may not find content that aligns with their interests, leading to reduced engagement and satisfaction. For each new post \( p_i \), a random embedding \( \mathbf{e}_{\textit{random}, i} \) is assigned:
    \[ \mathbf{e}_{\textit{random}, i} \sim \mathcal{N}(0, \sigma^2) \]

\subsubsection{Genre-based Average} In this method, we use a semi-personalized strategy by leveraging the average embeddings of high-view posts (\(v(c) \geq 10k)\) based on content genres. When new content is created within a specific genre, its embedding is initialized as the average embedding of high-view posts in that genre. This method offers a moderate level of personalization by considering the performance of previously successful content within the same genre, which enhances the likelihood of higher views for the new post. Let  \( G_j \)  represent the set of posts in genre  \( j \) , and  \( \mathbf{e}_{\textit{genre-avg}, j} \)  be the average embedding of high-view posts in that genre. For a new post  \( p_i \)  in genre  \( j \) , its embedding  \( \mathbf{e}_{\textit{genre-based}, i} \)  is initialized as the average genre embedding:

\begin{equation}
      \mathbf{e}_{\textit{genre-based}, i} = \mathbf{e}_{\textit{genre-avg}, j}
  \end{equation}  

where
\[ \mathbf{e}_{\textit{genre-avg}, j} = \frac{1}{|G_j|} \sum_{p \in G_j} \mathbf{e}_{p} \]

\subsubsection{Model-based} This involves a fully personalized model-based embedding initialization. Here, we develop a sophisticated model based on MEMER \cite{memer} that learns to initialize embeddings for new posts using both post embeddings of high-view content and multimodal content features (e.g., visual, text, audio). Details of the model are covered in the paper linked. Utilizing a model-based approach achieves a high degree of personalisation as the embeddings are tailored to individual posts, considering not only the genre but also the content's unique characteristics. This personalized initialization aims to enhance recommendations, leading to increased \textit{CVP}, \textit{CSR} and user satisfaction. The MEMER model is trained using interactions, high-view post embeddings (\( \mathbf{e}_{p} \)), user embeddings (\( \mathbf{u}_{p} \)), and multimodal content features (\( \mathbf{f}_{p} \)). However, during inference for a new post \( p_i \), only the content features (\( \mathbf{f}_{i} \)) can be utilized to infer a new embedding:
\[ \mathbf{e}_{\textit{model-based}, i} = \textit{MEMER\_Inference}(\mathbf{f}_{i}, \textit{MEMER}(\mathbf{e}_{p}, \mathbf{u}_{p}, \mathbf{f}_{p}, interactions)) \]

\noindent Summarising, selecting the appropriate embedding initialization has significant impact on how the $views_{min}$ are provided and how the content continues to progress beyond that. We look at offline metrics AUC, F1, RelaImpr \cite{yan2014coupled} for comparison of algorithms and online metrics to track the content progression (CVP) and user satisfaction (engagement).

\subsection{UI Surface Choices}
\label{ui_surface}
\begin{figure}[htb]
    \centering
    \includegraphics[width=0.45\textwidth]{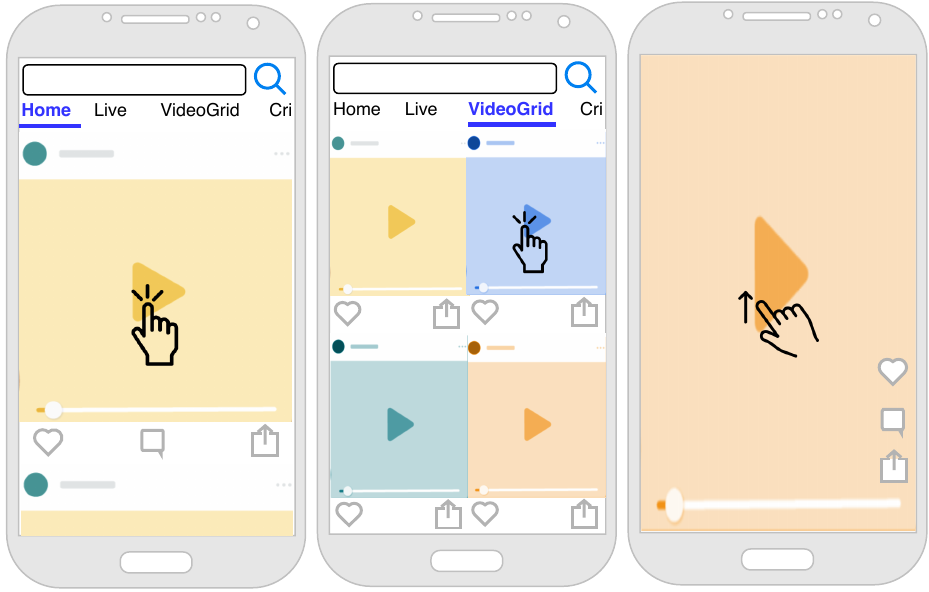}
    \caption{Representation of different UI Surface Choices - Home, VideoGrid, VideoScroll Feed resp.}
    \label{fig:screenshot}
    \vspace{-2pt}
\end{figure}

The user interface (UI) in a recommender system profoundly influences how users engage with the provided content. In this paper, we examine the effects of three UI surface designs on content progression and user interaction as illustrated in Figure \ref{fig:screenshot}.

\begin{enumerate}
    \item \textit{HomeFeed - Curated Choice}: It presents a vertically scrollable interface with videos that require a click to play. At any given time, a maximum of two videos are visible on the screen, similar to Instagram’s home feed. In a single feed fetch, around 12 videos are loaded, and the content refreshes as you scroll down. Users actively engage with the content, explicitly selecting videos based on their preferences. The intent is to encourage active user engagement and promote user-driven content selection.
    
    % \mohit{Include intent + business objective}
    \item \textit{VideoGridFeed - Variety Grid}: It displays video content as multiple tiles on a single screen, with each video requiring a click to play. The videos are arranged in a 4x4 grid, allowing users to view a variety of options at the same time. This setup aims to balance user choice and exposure, potentially leading to a more diverse content experience.
    
    \item \textit{VideoScrollFeed - Seamless Stream}: This UI offers autoplayed videos, and users can scroll to move to the next video, similar to popular platforms like Reels or TikTok. At any given time, only one video is in full-screen mode. On this surface, users consume content automatically without needing to make an explicit selection, so content progression is continuous and smooth. The goal is to make content consumption effortless and maximize consumption efficiency.
    % \mohit{Include intent + business objective}
\end{enumerate}

We investigate the impact of providing $views_{min}$ across different UI surfaces by measuring the CVP and engagement across them.

\begin{figure}[htb]
    \centering
    \includegraphics[width=8cm, height=5cm]{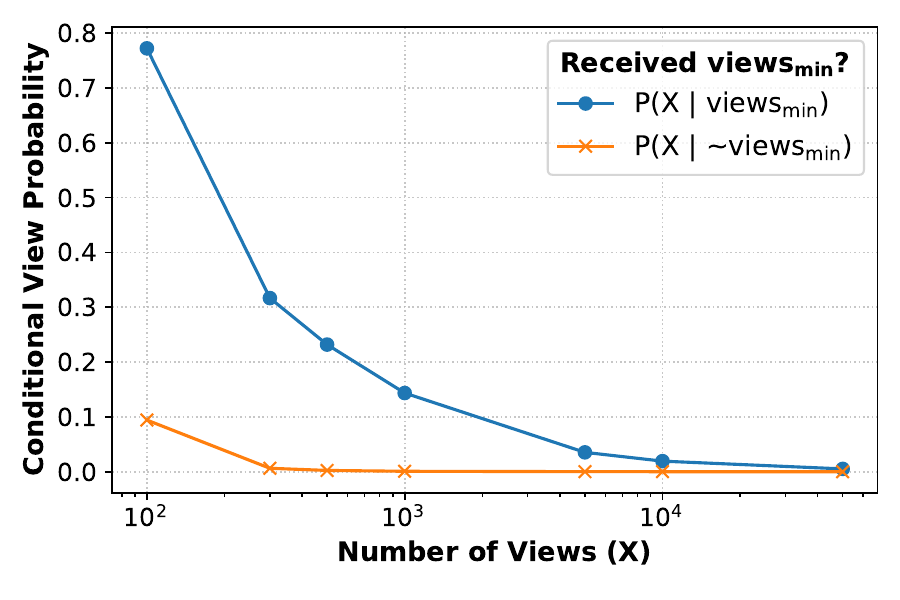}
    \caption{CVP if content receives $views_{min}$ \textit{vs.} it doesn't}
    % \caption{(\text{CVP} \mid v(c) \geq $views_{min}$) \textit{vs.} (\text{CVP} \mid v(c) \leq $views_{min}$)}
    \label{fig:cvp_given_mv}
\end{figure}

\begin{table*}[!htbp]
\centering
% \hspace{-10em}
\begin{minipage}[b]{0.32\linewidth}
  \centering
  \centerline{\begin{tabular}{|c|c|c|c|} \hline 
         &  \textbf{AUC}&  \textbf{F1}& \textbf{RelaImpr (\%)}\\ \hline 
         \textbf{Random}&  0.500&  0.151& -0.15\\ \hline 
         \textbf{Genre Average}&  0.605&  0.190& 66.7\\ \hline 
         \textbf{MEMER}&  0.631&  0.207& 83.3\\ \hline
    \end{tabular}}
    \subcaption{Offline Metrics for Algorithmic Choices for Successful Video Watch}
    \label{tab:offline_algo}
\end{minipage}
\hspace{10.5em}%
\begin{minipage}[b]{0.32\linewidth}
  \centering
  \centerline{\begin{tabular}{|c|c|c|} \hline 
         &  \textbf{Genre Average}& \textbf{MEMER}\\ \hline 
         $\textbf{CVP$(500|views_{min})$}$&  0.7235& 0.7646\\ \hline 
         \textbf{Engagement/Views}&  0.0099& 0.0145\\ \hline 
         \textbf{Successful Video Play}&  0.2633& 0.2871\\\hline
    \end{tabular}}
    \subcaption{Online metrics for Algorithmic Choices for User-Post AB Test}
    \label{tab:online_algo}
\end{minipage}
\vspace{-3pt}
\caption{Offline and Online Results from the Algorithmic Personalisation Choices}
\label{tab:algo choice tables}
% \vspace{-10pt}
\end{table*}

\begin{figure*}[!htbp]
\centering
% \hspace{-10em}
\begin{minipage}[b]{0.32\linewidth}
  \centering
  \centerline{\includegraphics[width=8cm, height=5cm]{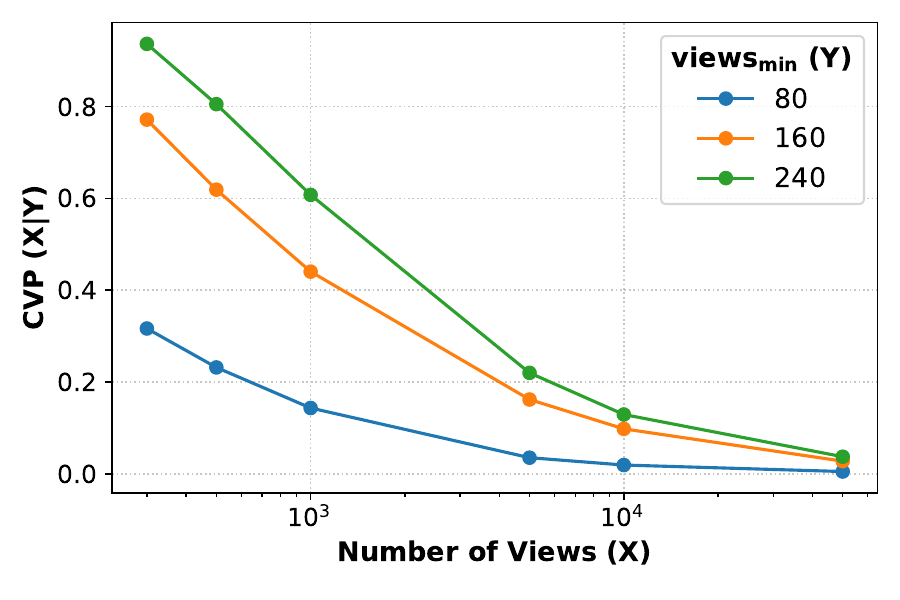}}
  \vspace{-5pt}
  \subcaption{CVP$(x|y)$}
  \label{fig:cvp_diff_mv}
\end{minipage}
\hspace{12em}%
\begin{minipage}[b]{0.32\linewidth}
  \centering
  \centerline{\includegraphics[width=8cm, height=5cm]{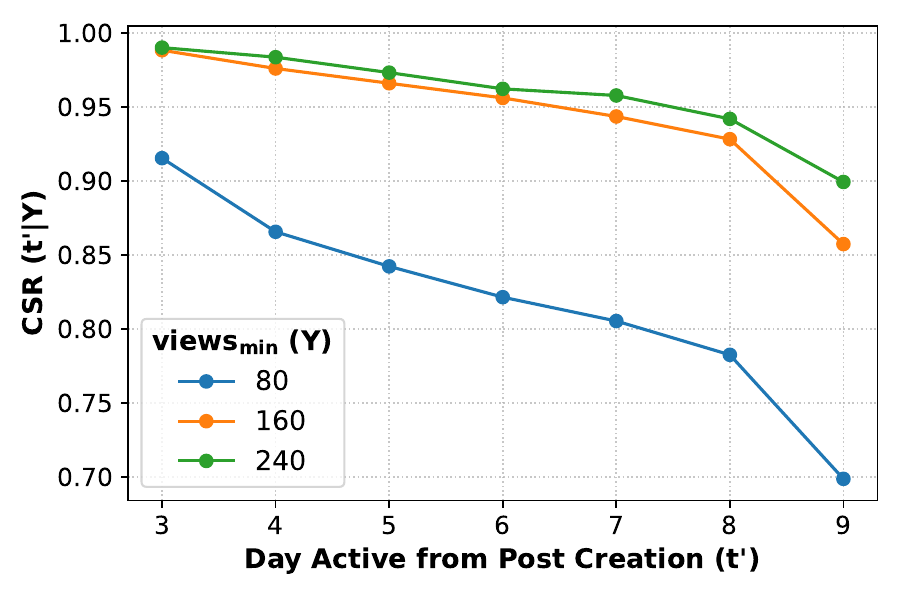}}
  \vspace{-5pt}
  \subcaption{CSR$(t'|y)$}
  \label{fig:csr_diff_mv}
\end{minipage}
\vspace{-3pt}
\caption{Content Progression Metrics given Differential $views_{min}$}
\label{fig:diff_minview}
% \vspace{-5pt}
\end{figure*}

\section{Experimentation \& Results}

\subsection{Experimentation Setup}

% \begin{figure*}[htbp]
% \vspace{-6pt}
% \centering
% % \hspace{-1em}
% \begin{minipage}[b]{0.4\linewidth}
%   \centering
%   \centerline{\includegraphics[width=14cm, height=2.2cm]{figures/experimentation.png}}
%   \vspace{-5pt}
%   % \subcaption{CVP if content receives $views_{min}$ vs not}
%   % \label{fig:cvp_ui}
% \end{minipage}
% \vspace{-3pt}
% \caption{The figure shows evolution of experimentation framework to gauge the impact of design choices across user satisfaction and platform content distribution (a) User level AB (b) User-Content AB (c) Parallel Experimentation Setup}
% \label{fig:experimentation}
% \vspace{-6pt}
% \end{figure*}
\subsubsection{User level AB}: In the conventional AB test setup, the user base is divided into two equal groups: test and control as shown in Fig \ref{fig:challenge_1}. The test group is exposed to treatment $\tilde{\textit{X}_{1}}$, while the control group is exposed to $\tilde{\textit{X}_{2}}$. Comparative metrics are used to measure the relative difference between the two groups, and the results are reported.

However, a major challenge arises due to the variable budget. For instance, if the test variant starts favoring highly-engaging categories like "Romance \& Relationships", the distribution of views across test and control groups becomes uneven. When this is scaled to 100\%, the relative numbers no longer hold true.

\subsubsection{User-Content AB}: This modification extends the previous setup by not only splitting the user base but also the content corpus; shown in Fig \ref{fig:challenge_2}. Each user set is paired 1:1 with a specific content set. This helps control spillage, as the effect of content set $\tilde{\textit{C}_{1}}$ is confined to user set $\tilde{\textit{U}_{1}}$ and does not impact users in the alternate set $\tilde{\textit{U}_{2}}$, and vice versa.

However, a limitation arises when experiments are confined to early-stage recommendation analysis. As the content progresses through its lifecycle, both groups may begin to influence each other, potentially leading to discrepancies in the scaled setup.

\subsubsection{Parallel Experimentation Setup}: This represents an advanced iteration of the user-content level AB testing. Here, the split is implemented across various stages of the content lifecycle, allowing for a more comprehensive evaluation. The framework is described in Fig \ref{fig:challenge_3}.
    
This setup is budget-conscious and enables the measurement of metrics beyond the early stage, such as assessing the impact of design choices during the cold-start phase on content progression up to expiration. Nevertheless, it is acknowledged that this setup is more intricate and costly to establish compared to the previous two methods. Additionally, there may still be minor data leakages if there are multiple sources contributing to views.

\subsection{Across System Design Choices}
% To setup the motivation initially for providing $views_{min}$, we compare two scenarios: one where content doesn't receive the promised views within 48 hours (can be experimented with) of creation, and the other. We note in Figure \ref{fig:cvp_given_mv} that content that attains the minimum pledged views within this time frame exhibits a higher progression rate (CVP) compared to those that do not. Notably, this difference is particularly pronounced for lower view thresholds ($<1k$) and eventually levels off.
\subsubsection{Impact of Meeting Minimum View Thresholds}
To establish the initial motivation for providing $views_{min}$, we compare two scenarios: one where content does not receive the promised views within 48 hours of creation (this time frame can be adjusted), and the other where it does. As shown in Figure \ref{fig:cvp_given_mv}, we observe that content that achieves the minimum pledged views within this time frame exhibits a higher progression rate (CVP) compared to content that does not. Notably, this difference is particularly pronounced for lower view thresholds ($<1k$) and eventually levels off.

\begin{figure}[htb]
    \centering
    \includegraphics[width=8cm, height=5cm]{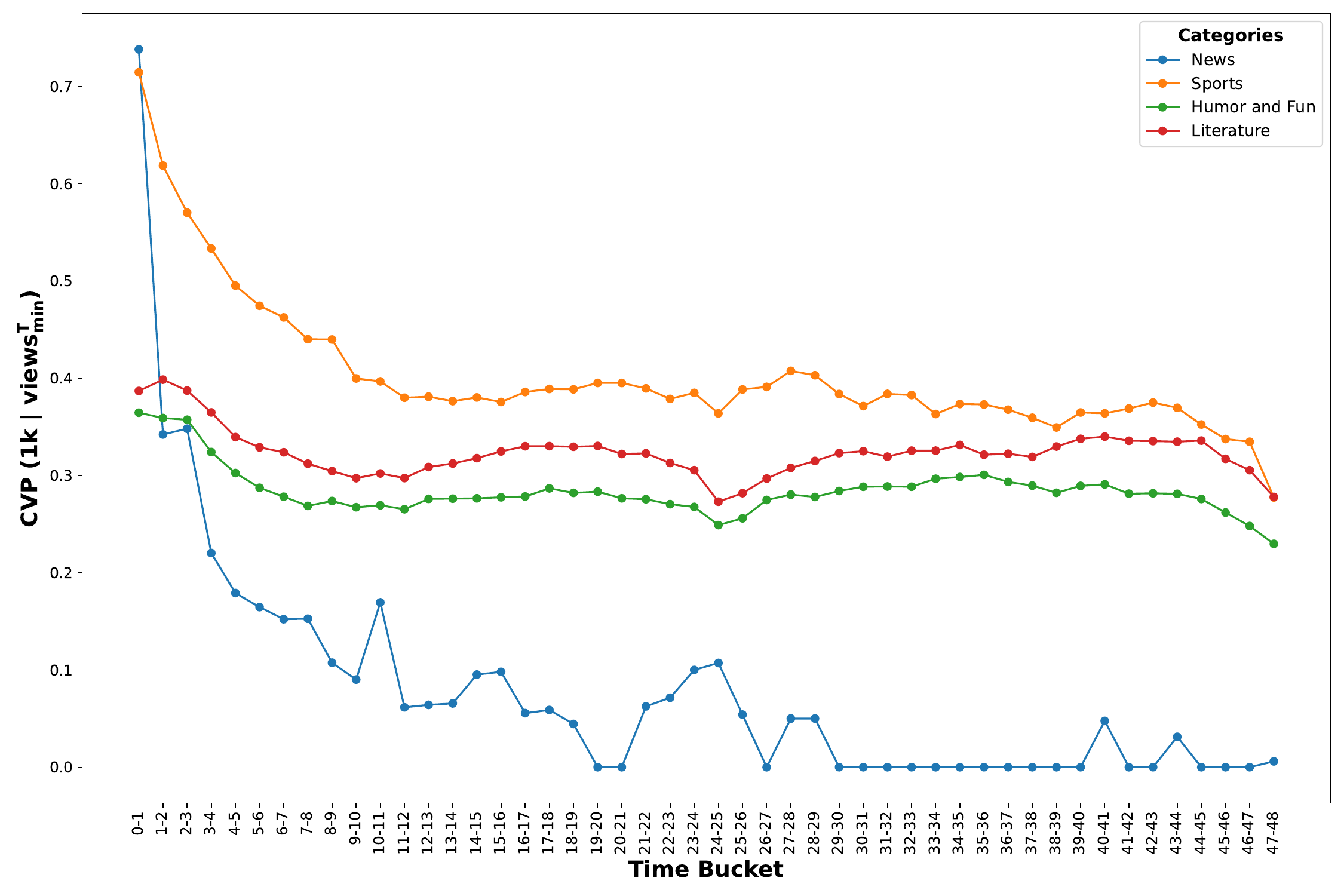}
    \caption{CVP($1k | views_{min}^{T}$) across content categories}
    \label{fig:cvp_cats}
\end{figure}

\begin{figure*}[htbp]
\centering
% \hspace{-10em}
\begin{minipage}[b]{0.32\linewidth}
  \centering
  \centerline{\includegraphics[width=8cm, height=5cm]{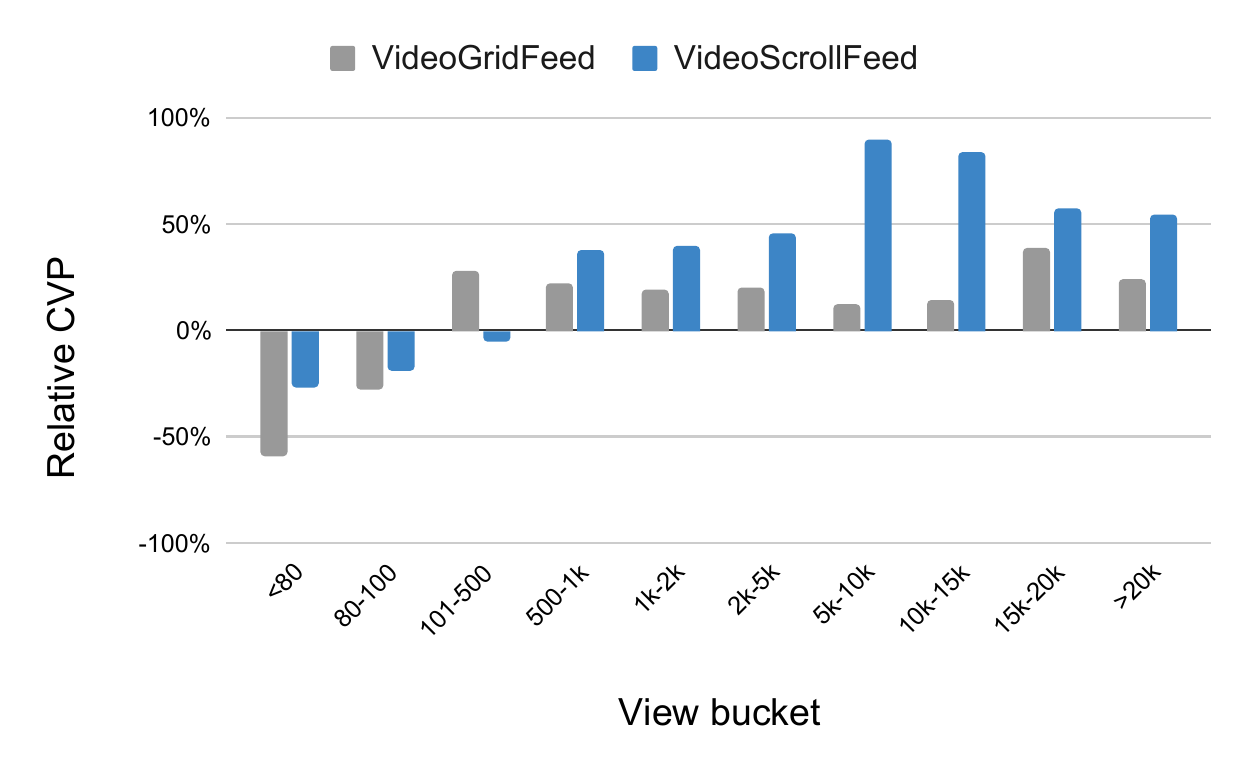}}
  \vspace{-5pt}
  \subcaption{Relative CVP across UI Surfaces}
  \label{fig:cvp_ui}
\end{minipage}
\hspace{11em}%
\begin{minipage}[b]{0.32\linewidth}
  \centering
  \centerline{\includegraphics[width=8cm, height=5cm]{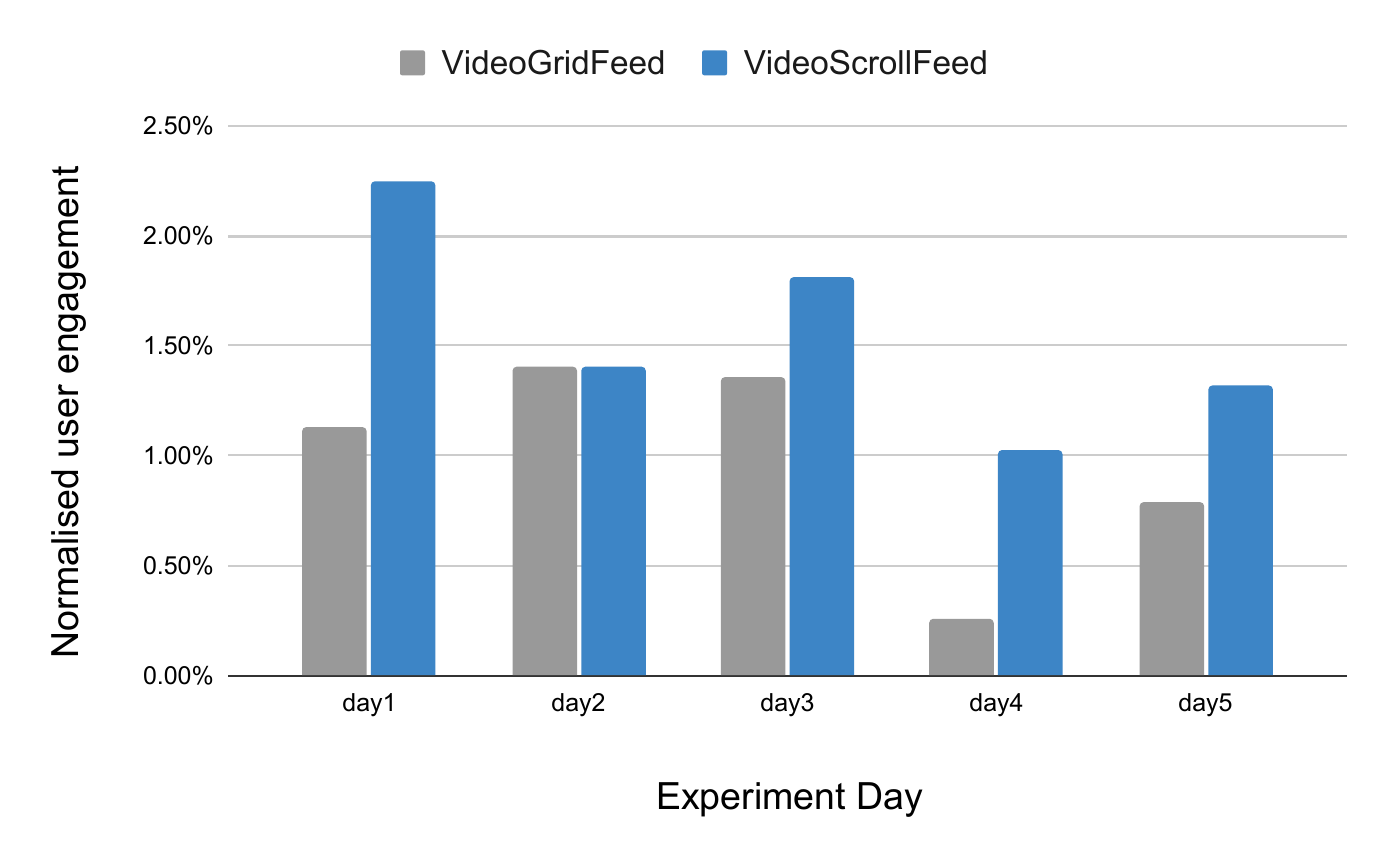}}
  \vspace{-5pt}
  \subcaption{User engagement \textit{w.r.t} Control}
  \label{fig:user_eng_ui}
\end{minipage}
\vspace{-3pt}
\caption{Content \& User Engagement Metrics across different UI surfaces}
\label{fig:diff_ui}
% \vspace{-10pt}
\end{figure*}

\subsubsection{Optimizing View Allocation for New Content}
Next, we explore varying the allocated view budget ($views_{min}$) for new content on the platform while keeping the algorithm and UI consistent throughout the experiment. In Figure \ref{fig:diff_minview}, we observe that generally, allocating a higher $views_{min}$ for content results in more posts achieving a greater number of views (CVP) and a longer shelf life (CSR). However, the relative increase becomes less pronounced as we continue to raise the $views_{min}$ threshold.

While this observation seems straightforward, its practical implications are more complex. It suggests that simply increasing $views_{min}$ won’t continue to positively affect content progression metrics indefinitely. This is because new content generally performs worse on the platform compared to \textit{mature} content due to the lack of user feedback data. This introduces an optimization challenge: we aim to maximize user satisfaction and content progression while ensuring that every new piece of content receives $views_{min}$. Without setting a cap on the exposure of new content, user satisfaction could be compromised, negatively impacting overall business metrics.

% Next, delving deeper, we vary the allocated view budget ($views_{min}$) itself for new content on the platform while maintaining a consistent algorithm and UI interface for the experiment. In Figure \ref{fig:diff_minview}, we observe that generally allocating a higher $views_{min}$ for content resulted in more posts achieving a greater number of views (CVP) and shelf life (CSR). However, the relative increase becomes less pronounced as we continue to increase the $views_{min}$ threshold.  
% While this observation is straightforward, its practical implication is not as simple. It suggests that simply increasing the $views_{min}$ wouldn't keep on positively affecting the content progression metrics. This is because the overall platform performance of new content tends to be less compared to \textit{mature} content due to the absence of user feedback data. Thereby, this introduces an optimization challenge where we aim to maximize user satisfaction and content progression while ensuring every new content receives $views_{min}$. Without setting a cap on the exposure of new content, user satisfaction could be compromised, impacting overall business metrics negatively.

\subsubsection{Analysis of View Rates Across Content Categories}
Finally, we conduct a more detailed analysis, exploring the impact of view rates across time-sensitive and timeless content categories. As shown in Figure \ref{fig:cvp_cats}, we find that for time-sensitive categories like \textit{News}, the Conditional-View-Probability (CVP) rate is greatly influenced by how quickly we provide the $views_{min}$. In fact, the drop in numbers varies significantly between the initial time buckets and later ones. This trend differs in categories like \textit{Humor \& Fun} videos, which are not time-sensitive. Posts in these categories tend to consistently achieve the target view count regardless of when the $views_{min}$ is provided. Please note that although there are 25 categories in the dataset, we have picked a few distinct ones to highlight the stark differences.

% Finally, we conduct a more detailed analysis, delving into the impact of view rates across time-sensitive and timeless content categories. From Figure \ref{fig:cvp_cats}, we find that for time-sensitive categories like \textit{News}, the Conditional-View-Probability (CVP) rate is greatly influenced by how promptly we provide the $views_{min}$. In fact, the drop in numbers varies significantly between initial time buckets and later ones. This trend differs in categories like \textit{Humor \& Fun} videos, which are not time-sensitive. Posts in these categories demonstrated a more consistent tendency to achieve the target view count irrespective of the timing of the $views_{min}$.

% For time sensitive genres e.g. news, minimizing latency is critical for achieving the desired view count. Our analysis demonstrated that when the $views_{min}$ were provided early in the post lifecycle, the post's tendency to reach the target view count was notably higher. In contrast, for non-time sensitive genres e.g. humor, the impact of latency on reaching the desired view count was less pronounced. 

\subsection{Across Algorithmic Personalisation Choices}
Initially, for motivation, Figure \ref{fig:personalize} shows the impact of varying personalisation on CVP. Next, we examine the effect of changing the underlying algorithm while keeping the initial view budget and UI surface constant for the experiment, assessing the impact on content progression and user satisfaction. We introduce three approaches: random allocation (for benchmarking), genre average, and MEMER, which are explained in Section \ref{algo_choices}. 
% Genre average uses embedding-based methods, deriving content embeddings with Field-aware-Factorization-Machine (FFM). For new content, it calculates an average within the category. MEMER, in contrast, employs meta-learning and multi-modal content embeddings to generate embeddings for new content.

% The offline and online results of different algorithmic choices are summarized in Table \ref{tab:algo choice tables}. We observe a notable improvement in offline AUC when using a sophisticated algorithm like MEMER compared to heuristic approaches such as random allocation or genre average (83.3\% \textit{RelaImpr} (RI) \cite{yan2014coupled} in Successful Video Watch [\ref{tab:offline_algo}]). These offline gains also translate into online metrics, reflected in higher CVP rates and increased user engagement \ref{tab:online_algo}. User engagement includes explicit signals such as likes, shares, and downloads, as well as implicit signals like successful video watch, which is a binary signal based on video watch time \ref{data_details}. 

The offline and online results of different algorithmic choices are summarized in Table \ref{tab:algo choice tables}. We observe a significant improvement in offline AUC when using a sophisticated algorithm like MEMER compared to heuristic approaches such as random allocation or genre average (83.3\% \textit{RelaImpr} (RI) \cite{yan2014coupled} for Successful Video Watch [\ref{tab:offline_algo}]). These offline gains also translate into online metrics, as seen in higher CVP rates and increased user engagement in Table \ref{tab:online_algo}. User engagement includes explicit signals such as likes, shares, and downloads, as well as implicit signals like successful video watch, which is a binary signal based on video watch time (explained in Sec \ref{data_details}).

% \begin{table}
%     \centering
%     \begin{tabular}{|c|c|c|c|} \hline 
%          &  \textbf{AUC}&  \textbf{F1}& \textbf{RelaImpr (\%)}\\ \hline 
%          \textbf{Random}&  0.500&  0.151& -0.15\\ \hline 
%          \textbf{Genre Average}&  0.605&  0.190& 66.7\\ \hline 
%          \textbf{MEMER}&  0.631&  0.207& 83.3\\ \hline
%     \end{tabular}
%     \caption{Offline Metrics for Algorithmic Choices for Successful Video Watch}
%     \label{tab:offline_algo}
% \end{table}

% \begin{table}
%     \centering
%     \begin{tabular}{|c|c|c|} \hline 
%          &  \textbf{Genre Average}& \textbf{MEMER}\\ \hline 
%          $\textbf{CVP$(500|views_{min})$}$&  0.7235& 0.7646\\ \hline 
%          \textbf{Engagement/Views}&  0.0099& 0.0145\\ \hline 
%          \textbf{Successful Video Play}&  0.2633& 0.2871\\\hline
%     \end{tabular}
%     \caption{Online metrics for Algorithmic Choices for User-Post AB Test}
%     \label{tab:online_algo}
% \end{table}

\subsection{Across UI Surfaces}
We keep the algorithm constant and introduce initial exposure to new content across various UI interfaces, as shown in Figure \ref{fig:screenshot}. We evaluate content performance across different view buckets and find that video feeds show higher Conditional-View-Probability (CVP) and user engagement rates compared to the control group (\textit{HomeFeed}), as depicted in Figures \ref{fig:cvp_ui} and \ref{fig:user_eng_ui}. Additionally, among video feeds, \textit{VideoScrollFeed} displays a higher CVP and user engagement rate, potentially because user intent is already captured at this stage of the funnel, making users more likely to explore fresh content. In contrast, \textit{VideoGridFeed} may show lower performance due to the added complexity of requiring a click on the thumbnail for the video to start playing, which impacts content progression metrics. The hierarchy of valuable impressions follows the order: \textit{VideoScrollFeed} $>$ \textit{VideoGridFeed} $>$ \textit{HomeFeed}.

% We keep the algorithm constant and introduce initial exposure to new content across various UI interfaces, description shown in Figure \ref{fig:screenshot}. We evaluate content advancement across different view buckets and find that video feeds exhibit superior Conditional-View-Probability (CVP) and User engagement rates compared to the control group (\textit{HomeFeed}), as depicted in Figure \ref{fig:cvp_ui} and \ref{fig:user_eng_ui}. Moreover, among video feeds, \textit{VideoScrollFeed} displays a higher CVP and user engagement rate, potentially due to user intent being already captured in this stage of the funnel. Users are thus more inclined to explore fresh content. Conversely, \textit{VideoGridFeed} may show lower performance due to the added complexity of clicking involved. Users must click on the thumbnail for the video to start playing, impacting content progression metrics. The hierarchy of valuable impressions follows the order: \textit{VideoScrollFeed} $>$ \textit{VideoGridFeed} $>$ \textit{HomeFeed}.

\section{Discussion and Conclusion}
We believe that design decisions have a significant impact on a system’s functionality. Our research explores the effects of these choices on content, examining various aspects, including system configurations, algorithmic selections, and UI designs. We show the tangible effects on content progression and longevity, and further analyze the differences in minimal exposure rates across different content categories. Additionally, we highlight the limitations of conventional A/B testing in accurately capturing content lifecycle metrics and propose more effective evaluation alternatives. This emphasizes the critical importance of informed design decisions in shaping the performance and longevity of content within the system.

There are several promising directions for future research. First, considering a variable budget allocation for new content could be beneficial, as it may not be necessary to allocate the same initial budget for every piece of new content. This adjustment could potentially free up resources to be redirected towards more promising content. Additionally, the distinct behavior observed in time-sensitive categories compared to others warrants further investigation. Exploring strategies for pacing initial views to ensure content relevance presents an intriguing area of study. Moreover, identifying the time-sensitive nature of content through behavioral feedback could offer valuable insights. Finally, an interesting area for future work involves evaluating the impact of design, algorithm, and system choices on content creators. This could be achieved by implementing relevant A/B test frameworks, providing insights into how these choices influence creation patterns and creator incentives.

\bibliographystyle{ACM-Reference-Format}
\bibliography{references}


\end{document}